\documentclass[journal]{IEEEtran}

\IEEEoverridecommandlockouts   

\usepackage{caption}
\usepackage{subcaption}
\usepackage[utf8]{inputenc} 
\usepackage[T1]{fontenc}    
\usepackage{url}            
\usepackage{booktabs}       
\usepackage{amsfonts}       
\usepackage{graphicx}%
\usepackage{amsmath}
\usepackage{amsthm}
\usepackage{comment}
\usepackage{amssymb}
\usepackage{xcolor}
\usepackage{accents}
\usepackage{caption}
\usepackage{cite}
\usepackage{cite}
\usepackage{xcolor}
\usepackage{multirow}
\usepackage{siunitx}
\usepackage{amsmath,bm}
\usepackage{algorithm}
\usepackage{algpseudocode}
\usepackage{multirow}

\newtheorem{definition}{Definition}
\newtheorem{lemma}{Lemma}

\newtheorem{remark}{\textbf{Remark}}

\title{\LARGE \bf
Distributed Estimation for a 3-D Moving Target in Quaternion Space with Unknown Correlation
}
\author{
\thanks{{\tt\small}.}
}
\author{
Yizhi Zhou$^{*}$, Yufan Liu$^{*}$ and Xuan Wang
\thanks{Y. Zhou and X. Wang are with the Electrical and Computer Engineering, George Mason University,
         Fairfax, VA, USA.
         Y. Liu is with the Electrical Engineering and Computer Science, University of California, Berkeley, Berkeley, CA, USA.
         Corresponding author: {\tt\small  xwang64@gmu.edu}}%
\thanks{$^{*}$ The authors contributed equally as co-first authors.}%
}

\begin{document}
\maketitle
\thispagestyle{empty}
\pagestyle{empty}
\raggedbottom

\begin{abstract}
For distributed estimations in a sensor network, the consistency and accuracy of an estimator are greatly affected by the unknown correlations between individual estimates.
An inconsistent or too conservative estimate may degrade the estimation performance and even cause divergence of the estimator. Cooperative estimation methods based on Inverse Covariance Intersection (ICI) can utilize a network of sensors to provide a consistent and tight estimate of a target. In this paper, unlike most existing ICI-based estimators that only consider two-dimensional (2-D) target state estimation in the vector space, we address this problem in a 3-D environment by extending the ICI algorithm to the augmented quaternion space. In addition, the proposed algorithm is fully distributed, as each agent only uses the local information from itself and its communication neighbors, which is also robust to a time-varying communication topology. To evaluate the performance, we test the proposed algorithm in a camera network to track the pose of a target. Extensive Monte Carlo simulations have been performed to show the effectiveness of our approach.
\end{abstract}

\section{Introduction} 
Multi-agent state estimation problem has been explored for decades,  driven by its extensive applications in fields such as target tracking \cite{Pzhu2021}, autonomous vehicle control \cite{Yizhi2023}, sensor networks \cite{Zbu2022}, surveillance \cite{Wang2009}, etc. Compared with centralized approaches, which are vulnerable to individual node failures at the fusion center and require heavy communication resources, distributed estimation has drawn more attention due to its scalability, robustness, and efficiency. Among the wide range of applications of distributed estimation, one important problem is the cooperative tracking of a moving target. In this problem, agents within the sensor networks can observe the target's state during a certain time period, either directly or indirectly, in the presence of measurement noise. Each agent aims to derive an accurate state estimate of the target by fusing its own information and that of its communicating neighbors. Solving this problem is nontrivial where two major challenges are as follows: The first challenge is to fuse the information despite unknown correlations. This is caused by the fact that cross-correlation between local information from different nodes always exists due to common process noise \cite{Lig1997}, which, however, is intractable in a distributed estimation scheme. Another challenge relates to tracking in 3-D environments, which is harder than in 2-D environments. This is because, in 3-D environments, the target state contains a 3-D rotation generally represented by a rotation matrix or a quaternion \cite{trawny2005}, which doesn't belong to the vector space. This issue restricts the application of tracking algorithms to those that are solely effective within vector space. Although one can still use the Euler angle to represent a 3-D rotation, it suffers from the well-known Gimbal lock problem \cite{Eka2005}.

The iterated-based \cite{Hlink2014} and diffusion-based \cite{Pyi2023} schemes have been widely used to design a consistent distributed Extended Kalman Filter (EKF) to cope with the unknown correlation regarding the first challenge. Most of them rely on the Covariance Intersection (CI) \cite{Ara2001} algorithm, which computes a convex combination of local estimates from one-hop communication. CI is also extended to fuse a quaternion in \cite{Pzhu2020} for 3-D tracking. However, reference \cite{Liji2002} points out the conservativeness of the CI, meaning that the CI often provides too conservative fusion results and makes the estimated covariance significantly larger than the actual covariance. Reference \cite{Rein2015} extensively illustrates that conservativeness may degrade estimation performance, which motivates the development of more optimal fusion rules.

Aside from CI and its extensive studies, an alternative approach named Inverse Covariance Intersection (ICI)  has been proposed in \cite{NOACK2017} for fusing two estimates with unknown correlation. This approach generates a consistent estimate and alleviates the conservativeness of CI fusion. In \cite{Ajgl2020}, ICI has been extended to the fusion of multiple estimates.
Most ICI-based filters \cite{Noa2019, Sun2023, SunT2023} are derived from the Information Filter (IF) (the information form of the EKF) and require the computing of the information vector. As a consequence, all these algorithms only work in the vector space that has additive errors. The pose of a 3-D moving target can be expressed by an \textit{augmented quaternion} \cite{Qi2023}, which contains a quaternion for orientation representation and a translation vector that represents the 3-D position. One cannot directly compute the information vector for a quaternion due to the dimension mismatching between a quaternion and its corresponding covariance matrix \cite{trawny2005}. Thus, the application of ICI algorithms to the 3-D tracking problem has not yet been addressed, limiting its applications in many real-world scenarios where the target has a 3-D motion. 

\textit{Statement of contribution:}
This paper aims to design a distributed estimation scheme with consistency and less conservativeness for 3-D target tracking problems. We first demonstrate the reason why ICI algorithms are not directly applicable to \textit{augmented quaternion spaces}. 
Then, we address this issue by extending the ICI algorithm in the context of error-state EKF to fuse the error state and then use it to update the nominal state in augmented quaternion.
The proposed framework is fully distributed, allows for time-varying communication topology among cameras, and is applicable to generic target motion and observation models. Finally, we perform extensive simulated experiments, using Monte-Corlo simulations in a camera sensor network to track a 3-D moving quadrotor, to validate the effectiveness of the proposed algorithm. The result of our algorithm shows the improvement in both accuracy and consistency over the CI-based approach, and is comparable to centralized fusion when cameras' communication probability is high.


\smallskip
\textit{Notations}:
Let $\bm I_{r\times r}$ denote the $r\times r$ identity matrix; $\mathbb{E}(\cdot)$ denote the expectation of a scalar; $\mathbf{Tr}(\cdot)$ denote the trace of a matrix. When applied to a set, $|\cdot|$ denotes the cardinality. 
We use $q\in\mathbb{R}^r$ to represent $q$ as a vector with all real entries and a dimension $r$. 
Let $\mathbb{G}^k(\mathcal{V},\mathcal{E}^k)$ be a directed graph representation of a time-varying sensor network, where $\mathcal{V}$ is the agent set and $\mathcal{E}^k$ is the edge set. If $(j, i)\in\mathcal{E}^k, j\neq i$, it means that agent $i$ can communicate with agent $j$ at time step $k$. The neighbor set is defined as $\mathcal{N}_i^k=\{j | (j,i)\in\mathcal{E}^k\}$. We always assume self-loop $(i,i)\in\mathcal{E}^k$ and $i\in\mathcal{N}_i^k$ for all $i\in\mathcal{V}$ and $k=1,2,\cdots$.

\section{preliminaries and Problem Formulation}

\subsection{Quaternion}
Quaternion is an efficient mathematical tool to represent rotation or orientation in 3-D space. According to the JPL~\cite{trawny2005} (v.s. Hamilton) definition, a quaternion can be expressed as: 
\begin{equation}
\mathbf{q} = q_1\mathbf{i} + q_2\mathbf{j} + q_3\mathbf{k} + q_4
\end{equation}
where \textbf{i}, \textbf{j}, and \textbf{k} are hyperimaginary numbers satisfying
\begin{align*}
\mathbf{i}^2 \!= \! \mathbf{j}^2 \! = \!\mathbf{k}^2  \!= \! -1,
-\mathbf{ij} = \mathbf{ji} = \mathbf{k}, -\mathbf{jk} = \mathbf{kj} = \mathbf{i}, -\mathbf{ki} = \mathbf{ik} = \mathbf{j}.
\end{align*}
Correspondingly, the vector form of $\mathbf{q}\in\mathbb{R}^4$ is\footnote{Although a more rigorous representing for quaternion space is $\mathbf{q}\in\mathbb{H}$, here, we use $\mathbb{R}^4$ to simply represent the dimension of the vector.}:
\begin{equation}
\mathbf{q}=\begin{bmatrix} 
\mathbf{q}_v \\ q_4\end{bmatrix}=\begin{bmatrix} 
q_1 &q_2 &q_3 &q_4\end{bmatrix}^\top.
\end{equation}
The $\mathbf{q}$ is a unit quaternion of rotation if it satisfies ${|\mathbf{q}|} = \sqrt{|{\mathbf{q}_v}|^2+q_4^2} = 1$, and can be further written as 
\begin{equation}
\mathbf{q} = \begin{bmatrix} \mathbf m sin(\theta/2) \\ cos(\theta/2) \end{bmatrix},
\end{equation}
where $\mathbf m\in\mathbb{R}^3$ is a unit vector that defines the rotation axis, and $\theta$ is the angle of rotation.
A rotation can also be described as a rotation matrix $\mathbf R \in \mathbb{R}^{3\times 3}$. The relation between a rotation matrix and a unit quaternion is
\begin{align}
\mathbf{R} = (2q_4^2-1)\bm{I}_3 - 2q_4\lfloor\mathbf q_v \times \rfloor + 2\mathbf q_v \mathbf q_v^\top,
\end{align}
where $\lfloor\mathbf q_v \times \rfloor$ is the skew-symmetric matrix~\cite{trawny2005}
\begin{align}
\lfloor\mathbf q_v \times \rfloor=\lfloor\begin{bmatrix}
    q_1\\q_2\\q_3
\end{bmatrix} \times \rfloor=\begin{bmatrix} 0 &-q_3 &q_2 \\ q_3 &0 &-q_1 \\ -q_2& q_1 &0\end{bmatrix}.
\end{align}
Based on the representation of a rotation, unit quaternions are also often used to represent orientation which specifies how an object should be rotated from a basis vector in 3D space to its current orientation.



Due to the constraints being a unit vector, quaternions representing rotations or orientations are not defined in a traditional vector space. As a consequence, one cannot use additive error to represent the error between two quaternions $\mathbf q_1$ and $\mathbf q_2$. Instead, the error is defined as a rotation $\delta{\mathbf{q}}=\begin{bmatrix}\delta \mathbf{q}_v^\top\quad \delta q_4\end{bmatrix}^\top$ between 
$\mathbf q_1$ and $\mathbf q_2$, which reads
\begin{align}\label{eq_err_q}
\delta \mathbf q=\mathbf{q}_1^{-1}\otimes\mathbf{q}_2=
\begin{bmatrix}
\Xi(\mathbf{q}_1)\quad \mathbf{q}_1
\end{bmatrix}^\top\mathbf{q}_2,
\end{align}
where $\otimes$ denotes the quaternion multiplication and the operator $\Xi(\cdot)$ for a given quaternion $\mathbf q$ is defined as:
\begin{align}\label{quat_err}
\Xi(\mathbf{q})=
\begin{bmatrix}
q_4 I_{3\times 3}-[\mathbf{q}_v\times]\\
-\mathbf{q}_v^\top
\end{bmatrix}
\end{align}.

If for error $\delta{\mathbf{q}}$, the associated rotation angle $\delta {\theta}$ is assumed to be sufficiently small, then $\delta{\mathbf{q}}$ can be approximated as
\begin{equation}\label{q_apprx}
\delta{\mathbf{q}} = \begin{bmatrix} 
\delta{\mathbf{q}_v} \\ \delta{q_4} \end{bmatrix} = \begin{bmatrix} 
\mathbf m sin(\delta{\theta/2}) \\ cos(\delta{\theta}/2) \end{bmatrix} \approx \begin{bmatrix} \frac{1}{2} \delta{\bm{\Theta}} \\ 1 \end{bmatrix},
\end{equation}
where $\delta{\bm\Theta}=\mathbf m\delta{\theta}\in\mathbb{R}^{3}$. If considering $\mathbf{\hat q}=\mathbf{q}_1$ and $\mathbf{q}^{\star}=\mathbf{q}_2$, respectively, as the estimated and true value of a quaternion, the uncertainty of $\delta{\mathbf{q}}$ is evaluated by the covariance matrix $\mathbf{P}\in\mathbb{R}^{3\times 3}$ of $\delta{\bm \Theta}$.

\subsection{Track-to-Track Fusion for Vector Space Variables}

Track-to-track fusion is the problem of combining multiple estimates of a state into a single estimate, which is more accurate and remains \textit{consistent}. 
\begin{definition}
    An estimate $(\mathbf{\hat x}, \mathbf{\hat P})$ of a state $\mathbf{\hat x}$ is consistent if $\mathbf{\hat P}\geq \mathbb E[(\mathbf{\hat x}-\mathbf x)^\top(\mathbf{\hat x}-\mathbf x)]$.
\end{definition}

To illustrate the technical challenges involved, consider a two-sensor fusion scenario with two \textit{consistent} estimates $(\mathbf{\hat x}_1, \mathbf{\hat P}_1)$ and $(\mathbf{\hat x}_2, \mathbf{\hat P}_2)$ of a state $\mathbf x$. 
We can compute a fused estimate $(\mathbf{\hat x}_f, \mathbf{\hat P}_f)$ by the most common linear combination approach
\begin{align}\label{eq_lc_x}
\mathbf{\hat x}_f&=K_1 \mathbf{\hat x}_1+K_2 \mathbf{\hat x}_2
\end{align}
with the fusion gains $K_1$ and $K_2$. The corresponding covariance matrix yields
\begin{align}\label{eq_lc_p}
\mathbf{\hat P}_f=K_1\mathbf{\hat P}_1 K_1^\top+K_2\mathbf{\hat P}_2 K_2^\top+K_1 \mathbf{\hat P}_{12} K_2^\top+K_2 \mathbf{\hat P}_{21} K_1^\top
\end{align}
where $\mathbf{\hat P}_{12}=\mathbf{\hat P}_{21}^\top$ are the cross-correlation terms. 
When the cross-correlations are known, one can compute $K_1$ and $K_2$ which guarantee the consistency of the fused estimate $(\mathbf{\hat x}_f, \mathbf{\hat P}_f)$ \cite{Juli1997}.
However, the correlation is generally non-zero, and in distributed sensor networks, it's unknown and intractable. Achieving a consistent fusion in such a case is nontrivial. To address this issue, the following fusion rules have been proposed to cope with the unknown correlations.  


\textit{Covariance Intersection (CI)}.
The well-known CI algorithm plays an important role in the fusion problem. 
Given two consistent local estimates $(\mathbf{\hat x}_1,\mathbf{\hat P}_1)$ and $(\mathbf{\hat x}_2,\mathbf{\hat P}_2)$, CI algorithm provides a consistent fused estimate $(\mathbf{\hat x}_f,\mathbf{\hat P}_f)$ given by
\begin{align}\label{eq_CI}
\mathbf{\hat P}_{CI}^{-1}&=\alpha(\mathbf{\hat P}_1)^{-1}+(1-\alpha)(\mathbf{\hat P}_2)^{-1},\quad \alpha\in[0,1]\nonumber\\
\mathbf{\hat P}_{CI}^{-1} \mathbf{\hat x}_{CI}&=\alpha(\mathbf{\hat P}_1)^{-1}\mathbf{\hat x}_1+(1-\alpha)(\mathbf{\hat P}_2)^{-1}\mathbf{\hat x}_2
\end{align}  
It is worth noting that the CI approach completely discards the cross-correlation terms (c.f. \eqref{eq_lc_p}), yielding an over-conservative result, i.e., the estimated covariance matrix can be much larger than the true error covariance matrix.

\textit{Inverse Covariance Intersection (ICI)}.
Unlike the CI rule, which ignores the correlation terms, the ICI algorithm seeks to derive the correlated components among the estimates $(\mathbf{\hat x}_1,\mathbf{\hat P}_1)$ and $(\mathbf{\hat x}_2,\mathbf{\hat P}_2)$. This yields a consistent estimate yet a tight covariance bound, which can substantially improve estimation accuracy. Specifically, ICI admits, 
\begin{align}
\mathbf{\hat P}_f^{-1} &= \mathbf{\hat P}_1^{-1}+\mathbf{\hat P}_2^{-1}-(\alpha \mathbf{\hat P}_1 + (1-\alpha) \mathbf{\hat P}_2)^{-1}\label{ICI_P}\\ 
\mathbf{\hat P}_f^{-1} \mathbf{\hat x}_f &= K_1 \mathbf{\hat x}_1 + K_2 \mathbf{\hat x}_2\label{ICI_x}
\end{align}
where $K$ and $L$ are the gain and can be computed by
\begin{align}\label{ICI_gain}
K_1&= \mathbf{\hat P}_1^{-1}-\alpha(\alpha \mathbf{\hat P}_1+(1-\alpha) \mathbf{\hat P}_2)^{-1}\nonumber\\
K_2&= \mathbf{\hat P}_2^{-1}-(1-\alpha)(\alpha \mathbf{\hat P}_1+(1-\alpha) \mathbf{\hat P}_2)^{-1}
\end{align}
While traditionally only applicable to two-sensor fusion \cite{NOACK2017},
ICI has recently been generalized~\cite{Ajgl2020} to fuse multiple estimates $(\mathbf{\hat x}_i,\mathbf{\hat P}_i),\,i=1,...,n$ as
\begin{align}\label{ICI_multi}
\mathbf{\hat P}_{ICI}&=
\begin{bmatrix}
\sum_{i=1}^n(\mathbf{\hat P}_i)^{-1}-(n-1)\mathbf {P_\beta}
\end{bmatrix}^{-1}\nonumber\\
\mathbf{\hat P}_{ICI}^{-1}\mathbf {\hat x}_{ICI}&=\sum_{i=1}^n
\begin{bmatrix}
(\mathbf{\hat P}_i)^{-1}-(n-1)\alpha_i \mathbf{P_\beta}
\end{bmatrix}\mathbf{\hat x}_i
\end{align}
where $\mathbf {P_\beta}=(\sum_{i=1}^n \alpha_i \mathbf{\hat P}_i)^{-1}$, and parameter $\alpha_i\in[0,1]$ satisfying $\sum_{i=1}^n\alpha_i=1$. The specific choices of $\alpha_i$ are specified in Lemma 7 of~\cite{Ajgl2020}.
\begin{lemma}\label{LM_lm1}\cite{NOACK2017}
   Let $\mathbf{\hat P}_{CI}^k$ and $\mathbf{\hat P}_{ICI}^k$ be, respectively, the fused covariance matrices with minimal traces using CI and ICI algorithms at timestep $k$. We have 
   $\mathbf{\hat P}_{CI}^k \geq \mathbf{\hat P}_{ICI}^k$.
\end{lemma}
    However, while the ICI method has seen significant application for variables in vector space and has been extended to fully distributed setups~\cite{SunT2023}, its formulation restricts its direct application to non-vector space variables. A direct observation of this limitation is that for quaternions, the information vector $\mathbf{P}^{-1}\mathbf{\hat q}$ suffers from a dimension mismatch with $\mathbf{P}\in\mathbb{R}^{3\times 3}$ and $\mathbf{\hat q} \in \mathbb{R}^4$.


\subsection{Problem Formulation: Cooperative 3-D Target Tracking with Quaternion States using Distributed Fusion}\label{Sec_PF}
Consider a moving target in the 3-D environment. Let ${G}$ and ${L}$ represent the global frame and the target’s local body frame, respectively.  The states of the target in 3-D are defined as:
\begin{align}
\mathbf x=
\begin{bmatrix}
^L_G \mathbf{q}^\top& ^G\mathbf{p}^\top&^G\mathbf{v}^\top
\end{bmatrix}^\top\in\mathbb{R}^{10}
\end{align}
which includes the agents' orientation described by quaternion $^{L}_{G} \mathbf{q}\in\mathbb{R}^4$ (the rotation quaternion from frame ${G}$ to ${L}$), 3-D position $^{G} \mathbf p \in\mathbb{R}^3$, and linear velocity $^{G}\mathbf v \in\mathbb{R}^3$ in the global frame. Consider that the target moves according to the following dynamics model (with a specific example in \eqref{IMU_dyn})
\begin{align}\label{eq_tgmdl}
\mathbf x^k=f(\mathbf x^{k-1},\mathbf{r}^{k-1})
\end{align}
where $\mathbf x^k$ is the target's state at time step $k$, and $\mathbf r^k\in\mathbb{R}^{r}$ is the zero mean white Gaussian noise with covariance $\mathbf Q\in\mathbb{R}^{r\times r}$. 

Now, suppose we have a sensor network $\mathbb{G}^k(\mathcal{V},\mathcal{E}^k)$ of $|\mathcal{V}|=n$ agents, where each agent knows target dynamics \eqref{eq_tgmdl} and can sense the target with a limited sensing region. The agents also have the ability to communicate, through the time-varying edges $\mathcal{E}^k$, with their neighbors.
At each time step $k$, if agent $i$ sees the target, it obtains a local measurement $z_i^k\in\mathbb{R}^m$ related to the target's state $\mathbf x^k$ following a nonlinear measurement model (with a specific example in \eqref{cam_mdl})
\begin{align}
z_i^k = h_i(\mathbf x^k)+\mathbf w_i^k
\end{align}
where $\mathbf w_i^k\in\mathbb{R}^m$ is the local measurement noise generally assumed to be zero-mean Gaussian noise with covariance $\mathbf R_i\in\mathbb{R}^{m\times m}$. 
Note that $\mathbf w_i^k$ and $\mathbf r^k$ are mutually independent, as they are generally generated by different devices. 

The objective of this paper is to obtain an accurate target state estimation $(\mathbf {\hat x}_i^k, \mathbf {\hat P}_i^k)$ for each agent, by only using its local information and the information received from its neighbor set $\mathcal{N}_i^k$.  The key challenge lies in handling the states represented by augmented quaternion and ensuring that the estimate is consistent. The second challenge arises from the fact that the correlation among agents estimates $(\mathbf {\hat x}_i^k, \mathbf {\hat P}_i^k)$ always exists since each agent has to process a common target motion model as equation \eqref{eq_tgmdl}, which is unknown and intractable in a distributed scheme\cite{Lig1997}.




\section{Proposed Algorithm}
To solve the formulated problem, a feasible solution using a CI-based method has been proposed in \cite{Cras2009}. However, as we have demonstrated the advantage of ICI over CI in Lemma \ref{LM_lm1}, our goal is to generalize the ICI-based method to achieve a fully distributed scheme for the cooperative estimation of a 3-D moving target with quaternion states. 
The proposed algorithm is as follows and is summarized in Algorithm \ref{alg}.

\medskip
\noindent \textbf{Propagation}: 
Suppose that at time-step $k$, each agent carries a posterior estimate $(\mathbf{\hat{x}}_i^{k-1}, \mathbf{\hat{P}}_i^{k-1})$ of the previous time-step. The first step is to propagate the posterior estimation pair $(\mathbf{\hat{x}}_i^{k-1}, \mathbf{\hat{P}}_i^{k-1})$ to obtain a prior estimation pair $(\mathbf{\bar{x}}_i^{k}, \mathbf{\bar{P}}_i^{k})$ based on the target motion model, which reads
\begin{align}\label{eq_pg1}
    \mathbf{\bar x}_i^k&=f({\mathbf{\hat x}_i^{k-1}})\nonumber\\
    \mathbf{\bar P}_i^k&=\Phi_i^{k-1} \mathbf{\hat P}_i^{k-1} (\Phi_i^{k-1})^\top+\mathbf{O}_i^{k-1}
\end{align}
where 
\begin{align}\label{eq_pg2}
    \Phi_i^{k-1}&=\frac{\partial f}{\partial \mathbf x}(\mathbf{\hat x}_i^{k-1}),\quad
    \mathbf{G}_i^{k-1}=\frac{\partial f}{\partial \mathbf r}(\mathbf{\hat x}_i^{k-1}),\quad\nonumber\\
    \mathbf{O}_i^{k-1}&=\mathbf{G}_i^{k-1}\mathbf{Q}(\mathbf{G}_i^{k-1})^\top,
\end{align}
and $\mathbf{Q}$ is the covariance matrix of the noise in~\eqref{eq_tgmdl}, which is known to all agents.  

\medskip
\noindent\textbf{Intermediate estimation}: 
In this step, each agent $i$ will first communicate with its neighbors to obtain the neighbors' prior estimates $(\bar{\mathbf{x}}_j^k, \bar{\mathbf{P}}_j^k)$ for all $j \in \mathcal{N}_i^k$. Then, the agent aims to update its local estimate $(\mathbf{\bar{x}}_i^{k}, \mathbf{\bar{P}}_i^{k})$ to a more accurate intermediate estimate $(\mathbf{\check{x}}_i^{k}, \mathbf{\check{P}}_i^{k})$ by fusing its local information with the information obtained from its neighbors. During this process, even if some agents cannot observe the target due to distance, they can still use their neighboring agents' estimates to correct their local estimate.

To facilitate the fusion, recall that in distributed settings, tracking the correlations is impossible. Although the ICI algorithm can fuse estimates with unknown correlations, its application has been limited to vector spaces through the computation of the information vector. It remains unclear how to fuse states represented by an augmented quaternion, which includes both a quaternion and a translation vector. To address this issue, we extend the ICI algorithm in the context of the error state EKF as follows.
Define the error state $\delta \bar{\mathbf x}_{ij}^k$ between the prior estimates $\mathbf{\bar x}_i^k$ of agent $i$ and that of its neighbor's $\mathbf{\bar x}_j^k$ at the time-step $k$ as:
\begin{align}\label{err_state}
    \delta  \bar{\mathbf x}_{ij}^k&=
    \begin{bmatrix}
    {\delta \bar{\bm{\Theta}}_{ij}^k}^{\top} & 
    {\delta \bar{\mathbf{p}}_{ij}^k}^{\top} &
    {\delta \bar{\mathbf{v}}_{ij}^k}^{\top}
    \end{bmatrix}^{\top}\in\mathbb{R}^9
\end{align}
where position error ${\delta\bar{\mathbf{p}}_{ij}^k} = ^G\mathbf{\bar{p}}_i^k-^G\mathbf{\bar{p}}_j^k \in\mathbb{R}^3$ and velocity error ${\delta \bar{\mathbf{v}}_{ij}^k}= ^G\mathbf{\bar v}_i^k-^G\mathbf{\bar v}_j^k \in\mathbb{R}^3$ are directly computed using the arithmetic difference. 
The error corresponding to the quaternion components $\mathbf{^L_G{\bar q}}_i^k$ and $\mathbf{^L_G{\bar q}}_j^k$, assuming the required rotation is small, is defined as ${\delta \bar{\bm{\Theta}}_{ij}^k}\in\mathbb{R}^3$, which is approximated from 
$$\Xi^\top(\mathbf{^L_G{\bar q}}_j^k)\mathbf{^L_G{\bar q}}_i^k\approx \frac{1}{2} {\delta \bar{\bm{\Theta}}_{ij}^k},$$
following the definitions \eqref{eq_err_q}-\eqref{q_apprx}.



    
The error definition in \eqref{err_state} resolves the dimension mismatch and allows us to generalize the multi-estimate ICI algorithm \cite{Ajgl2020} to fuse all error estimates $\delta \mathbf x_{ij}^k, j\in\mathcal{N}_i^k$ including quaternion variables to compute the fused covariance matrix $\check {\mathbf P}_i^k \in\mathbb{R}^{9\times 9}$
    \begin{align}\label{err_ICI_p}
    \check {\mathbf P}_i^k&=
    \begin{bmatrix}
    \sum_{j\in\mathcal{N}_i^k}(\bar {\mathbf P}_j^k)^{-1}-(|\mathcal{N}_i^k|-1) ({\mathbf P}_{\beta,i}^k)^{-1}
    \end{bmatrix}^{-1}\end{align}
    where
    \begin{align*}
    {\mathbf P}_{\beta,i}^k&=\sum_{j\in\mathcal{N}_i^k}\alpha_{ij}^k\bar{\mathbf P}_j^k
    \end{align*}
    and the fused state correction vector $\delta \bar{\mathbf x}_i^k$ is defined as 
    \begin{align}\label{err_ICI_x}
    \delta \bar{\mathbf x}_i^k
    &=\check {\mathbf P}_i^k \sum_{j\in\mathcal{N}_i^k}
    \left(
    (\bar {\mathbf P}_j^k)^{-1}-(|\mathcal{N}_i^k|-1)\alpha_{ij}^k {(\mathbf P}_{\beta,i}^k)^{-1}
    \right)\delta \bar{\mathbf x}_{ij}^k\nonumber\\
    &=
    \begin{bmatrix}
    {\delta \bar{\bm{\Theta}}_{i}^k}^{\top} & 
    {\delta \bar{\mathbf{p}}_{i}^k}^{\top} &
    {\delta \bar{\mathbf{v}}_{i}^k}^{\top}
    \end{bmatrix}^{\top}\in\mathbb{R}^9
    \end{align}
    which is partitioned in the same way as \eqref{err_state}.
    For the parameters $\alpha_{ij}^k$ in equations \eqref{err_ICI_p}-\eqref{err_ICI_x}, we refer to the following choice in \cite{Ajgl2020} to ensure that ICI provides a tight covariance bound.    
    \begin{align}\label{w_i}
    \alpha_{ij}^k=\frac{1/\textbf{Tr}(\mathbf{\bar P}_j^k)}{\sum_{j\in \mathcal{N}_i^k} 1/\textbf{Tr}(\mathbf{\bar P}_j^k)}
    \end{align} 

    To continue, we update the local prior estimation $(\mathbf{\bar x}_i^{k}, \mathbf{\bar P}_i^{k})$ using the fused error correction $\delta \bar{\mathbf x}_i^k$. This is enabled by the following $\boxplus$ operator:
    \begin{align}\label{eq_interup}
    \check {\mathbf x}_i^k&=\bar {\mathbf x}_i^k\boxplus\delta {\bar{\mathbf x}}_i^k=
    \begin{bmatrix}
        ^L_G\mathbf{\bar q}_i^k\otimes \delta \bar{\mathbf q}_i^k\\
        ^G\bar {\mathbf p}_{i}^k+{\delta \bar{\mathbf{p}}_{i}^k}\\
        ^G\bar {\mathbf v}_{i}^k+{\delta \bar{\mathbf{v}}_{i}^k}
    \end{bmatrix}
    \end{align}  
    where $\otimes$ is the quaternion multiplication and  
    \begin{align}\label{eq_convertq}
    \delta \bar{\mathbf q}_i^k\triangleq
    \frac{1}{\sqrt{1+\frac{1}{4}(\delta\bm\Theta_i^k)^\top (\delta\bm\Theta_i^k)}}
    \begin{bmatrix}\frac{1}{2}\delta\bm \Theta_i^k \\ 1\end{bmatrix}.
    \end{align}
    In \eqref{eq_interup}, the quaternion is updated by rotation. The positions and velocities are updated by direct addition. Equation \eqref{eq_convertq} converts the fused quaternion error $\delta\bm\Theta_i^k$ back to a rotation quaternion by re-scaling it to ensures that the produced $\delta \bar{\mathbf q}_i^k$ is a unit quaternion.

\medskip
\noindent\textbf{Update}: The last step is to update the intermediate estimate $(\mathbf{\check x}_i^k, \mathbf{\check P}_i^k)$ with all local measurements $z_j^k, j\in\mathcal{N}_i^k$. Note that if an agent is blind (i.e., does not observe the target) for this time step, then \eqref{eq_S}-\eqref{eq_y} are set to zero. Otherwise, 
we linearize $z_i^k$ relative to the current state estimate to compute the measurement Jacobian  $\mathbf H_i^k=\frac{\partial h^k_i}{\partial \mathbf{x}_i}(\mathbf{\check x}_i^k)\in\mathbb{R}^{m\times 9}$. 
Then we compute
    \begin{align}\label{eq_S}
        \mathbf I_i^k&=(\mathbf H_i^k)^\top (\mathbf R_i^k)^{-1}\mathbf H_i^k\\
        \mathbf y_i^k&=(\mathbf H_i^k)^\top (\mathbf R_i^k)^{-1}(z_i^k-h_i^k(\mathbf{\check x}_i^k)), \label{eq_y}
    \end{align}
and updated covariance matrix $\mathbf{\hat P}_i^k$ and correction $\Delta x_i^k$ as
    \begin{align}\label{eq_ph}
    \mathbf{\hat P}_i^k&=
    \begin{bmatrix}
    (\mathbf{\check P}_i^k)^{-1}+\sum_{j\in\mathcal{N}_i^k}\mathbf I_j^k    
    \end{bmatrix}^{-1}\\\label{eq_dx}
    \Delta \mathbf{x}_i^k&=
    \begin{bmatrix}
    \Delta \bm \Theta_i^k\\
    \Delta \mathbf p_{i}^k\\
    \Delta \mathbf v_{i}^k
    \end{bmatrix}=
    \mathbf {\hat P}_i^k \sum_{j\in \mathcal{N}_i^k} \mathbf y_j^k.
    \end{align}
    In this update, agent $i$ has two ways to obtain $\mathbf I_j^k$ and  $\mathbf y_j^k$, the first one is by a new round of communication after their neighbors finish computing \eqref{eq_S} and \eqref{eq_y}. Alternatively, assuming $h_j^k$, $\mathbf R_j^k$, and $z_j^k$ are also shared during the intermediate estimation step, agent $i$ can compute these quantities for their neighbors, saving a round of communication. 
    Finally, we can compute the posterior estimate of the time-step $k$ as
    \begin{align}\label{eq_xhat}
    \hat{\mathbf x}_i^k=\check {\mathbf x}_i^k\boxplus \Delta {\mathbf x}_i^k
    \end{align}
    by following the definition of $\boxplus$ in \eqref{eq_interup}.
\begin{algorithm}
\caption{Cooperative Estimation Algorithm for 3-D Target Tracking Implemented by Agent $i$ at Timestep $k$}\label{alg}
\begin{algorithmic}
\State\textbf{Propagation:} Compute the prior estimation pair $(\mathbf{\bar x}_i^k,\mathbf{\bar P}_i^k)$ using \eqref{eq_pg1} and \eqref{eq_pg2}.
\State\textbf{Intermediate estimation:} Communicate with its neighbor $j, \forall j\in\mathcal{N}_i^k$, to obtain $(\mathbf{\bar x}_j^k, \mathbf{\bar P}_j^k)$. Then compute the state correction $\delta \bar{\mathbf{x}}_i^k$ to get the intermediate estimate $(\mathbf{\check x}_i^k, \mathbf{\check P}_i^k)$ using \eqref{err_ICI_p}, \eqref{err_ICI_p}, and \eqref{eq_interup}.
\State\textbf{Update:} Update the intermediate estimate $(\mathbf{\check x}_i^k, \mathbf{\check P}_i^k)$ with measurements $z_j^k, j\in\mathcal{N}_i^k$ to compute the posterior estimate $(\mathbf{\hat x}_i^k, \mathbf{\hat P}_i^k)$
\end{algorithmic}
\end{algorithm}

\begin{remark} [Implementation and Effectiveness of Algorithm \ref{alg} under Time-varying Graphs with Blind Agents]

In terms of implementation, by revisiting Algorithm \ref{alg}, the propagation step does not rely on either communication or new measurements.
The intermediate estimation step is where agent communication becomes involved. It is inherently applicable to time-varying graphs in the sense that agents perform fused estimation updates when neighboring agents' information is available, or they update using their own local information otherwise. The update step requires new measurements from agents. When an agent $i$ is blind, we can equivalently assume that $\mathbf{R}_i^k=\infty$, i.e., infinite measurement uncertainties for $z_i^k$ when agent $i$ is not able to directly detect the target. This yields $\mathbf I_i^k=0$ and $\mathbf y_i^k=0$ in \eqref{eq_S}-\eqref{eq_y}. If agent $i$ and all its neighbors are blind for this time step, agent $i$ performs no updates in \eqref{eq_ph} and \eqref{eq_dx}, which is consistent with our description.

In terms of effectiveness, the stability and consistency properties of the algorithm in error states can then be established in the same way as in \cite{Ajgl2020}. For measurement accuracy, while Lemma \ref{LM_lm1} only establishes the advantage of ICI over CI regarding the bound of the covariance matrix, to verify the performance of the algorithm in experiments, we will apply the proposed algorithm to track a drone in simulation with a comparison to the CI-based filter \cite{Cras2009}. We will evaluate using the Root Mean Square Error (RMSE) and the Normalized Estimation Error Squared (NEES) \cite{Huang2011}, which are the most common criteria for evaluating accuracy and consistency.
\end{remark}

\section{Simulations}
\begin{figure}[h!]
    \centering
    \includegraphics[width=0.48\textwidth]{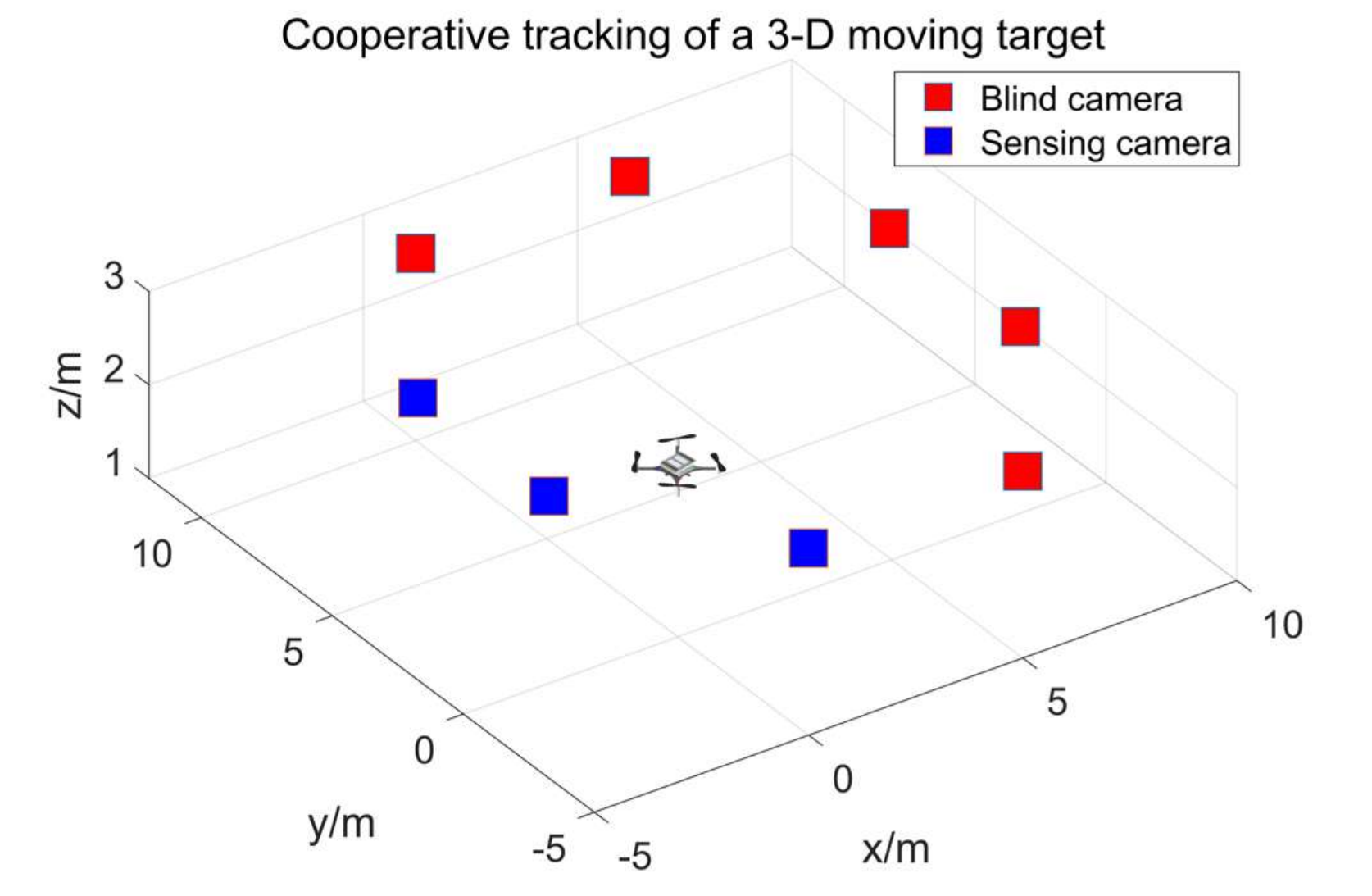}
    \caption{Cooperative tracking of a drone in 3-D environments over camera networks. The red cameras represent the blind cameras which are not able to sense the drone, while the blue ones indicate that the cameras can directly detect the drone.}
    \label{fig1}
\end{figure}

In this section, we employ the proposed ICI cooperative tracking algorithm over a camera network to validate its performance with extensive Monte-Carlo simulations, where 8 cameras are used to track the pose of a quadrotor that has a 3-D motion. In the simulation, we set that each camera only has a limited range of view which is 5 $m$, i.e., cameras are not able to detect the target when the quadrotor is out of the sensing range, as illustrated in Fig \ref{fig1}.
Additionally, we define \textit{communication rate} with certain percentages to quantify the connection of the cameras over the sensor networks. For example, $10\%$ means that each camera can communicate with every other camera with the probability of $10\%$ at each time step. Hence the communication graph over the sensor network is time-varying.

\medskip
\noindent \textbf{Simulation models:}
The proposed algorithm needs the target's dynamics model for state propagation.
One way to predict the target's state (position, orientation, and velocity) is by using data from an inertial measurement unit (IMU) for dynamics modeling. In this paper, we directly employ the discrete-time IMU propagation model as the target's motion model to obtain a prior estimate. For the measurement model, we apply the standard pinhole camera model to the camera networks, which comprise 8 cameras. The descriptions of the models can be found in the Appendix.

\medskip
\noindent \textbf{Simulation results:}
We let the quadrotor move following pre-designed trajectories and see if the proposed algorithm can accurately track them. Then, we perform 50 Monte-Carlo simulations and quantify the results using Root Mean Square Error (RMSE) to evaluate the accuracy and Normalized Estimation Error Squared (NEES) to show consistency.

\begin{table}[t]
\centering
\caption{Averaged RMSE of all the cameras over 50 Monte-Corlo simulations, compared with the results of the centralized approach.}
\begin{tabular}{*{4}{c}}
  \toprule
  \multirow{2}*{Proposed} &
  \multirow{2}*{communication} & \multicolumn{2}{c}{Tracking error}\\
  \cmidrule(lr){3-4}
  & & Position (m) & Orientation (deg) \\
  \toprule
  \multirow{4}*{Path $a$} 
   & 0\% & 17.860 & 15.830 \\
   &20\% & 0.068 & 11.160 \\
   &40\% & 0.054 & 10.092\\
   &60\% & 0.052 & 9.650 \\
   &80\% & 0.049 & 9.010\\
  \midrule
  \multirow{4}*{Path $b$} 
   & 0\% & 22.068 & 11.830 \\
   &20\% &0.059 & 5.751 \\
   &40\% &0.045 & 5.362 \\
   &60\% &0.035 & 5.244 \\
   &80\% &0.032 & 4.915 \\
   \bottomrule
   \toprule
  \multirow{2}*{CI-based} &
  \multirow{2}*{communication} & \multicolumn{2}{c}{Tracking error}\\
  \cmidrule(lr){3-4}
  & & Position (m) & Orientation (deg) \\
  \toprule
  \multirow{4}*{Path $a$} 
   & 0\% & 17.860 & 15.830 \\
   &20\% & 0.061 & 12.780 \\
   &40\% & 0.057 & 12.715\\
   &60\% & 0.056 & 12.104 \\
   &80\% & 0.052 & 11.081\\
  \midrule
  \multirow{4}*{Path $b$} 
   & 0\% & 22.068 & 11.830 \\
   &20\% &0.064 & 7.938 \\
   &40\% &0.046 & 8.362 \\
   &60\% &0.039 & 8.763 \\
   &80\% &0.037 & 7.952 \\
   \bottomrule
   \toprule
   {\multirow{2}*{Centralized}} & \multicolumn{2}{c}{Tracking error}\\
   \cmidrule(lr){2-3}
   & Position (m) & Orientation (deg) \\
   \midrule
   Path $a$ & 0.048 & 9.001 \\
   \midrule
   Path $b$ & 0.029 & 4.895 \\
   \bottomrule
   \end{tabular}
\label{Tab1}
\end{table}


\begin{figure}[ht]
	\centering
	\begin{subfigure}{0.46\textwidth}
		\centering
		\includegraphics[width=\textwidth]{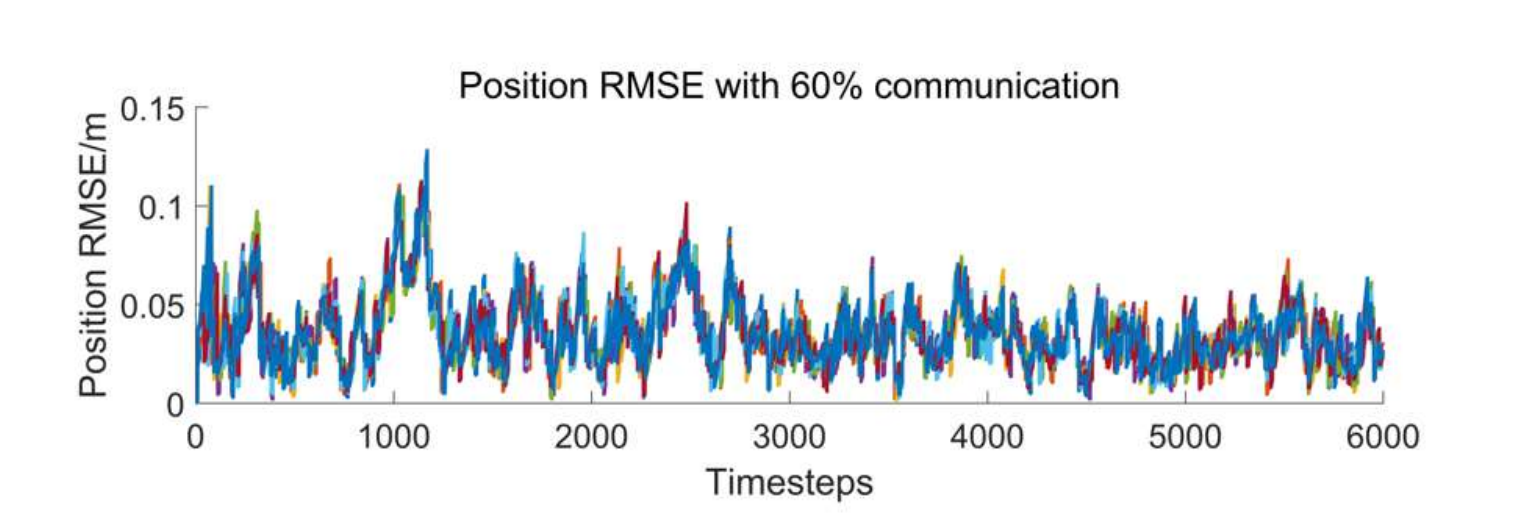}
		\caption{Position RMSE of each camera with $60\%$ communication rate}
	\end{subfigure}
        \begin{subfigure}{0.46\textwidth}
		\centering
		\includegraphics[width=\textwidth]{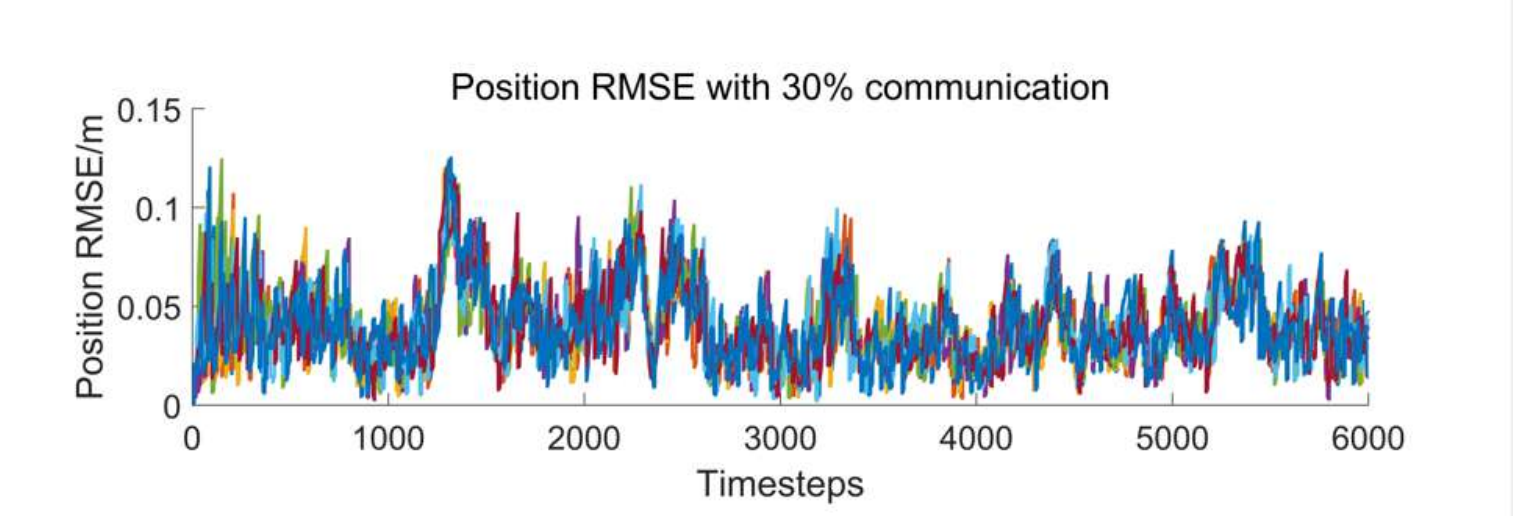}
		\caption{Position RMSE of each camera with $30\%$ communication rate}
	\end{subfigure}
        \caption{Position RMSE of each camera using the proposed algorithm with different communication rates.}
        \label{Fig_PRMSE}
\end{figure}

\begin{figure}[ht]
	\centering
	\begin{subfigure}{0.46\textwidth}
		\centering
		\includegraphics[width=\textwidth]{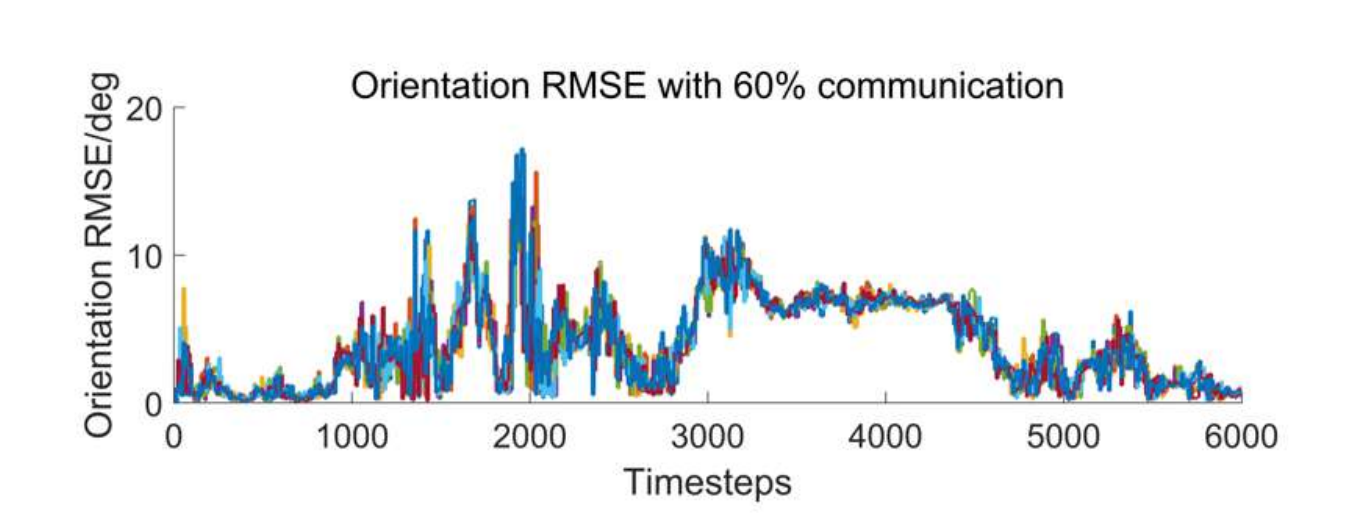}
		\caption{Orientation RMSE of each camera with $60\%$ communication rate}
	\end{subfigure}
        \begin{subfigure}{0.46\textwidth}
		\centering
		\includegraphics[width=\textwidth]{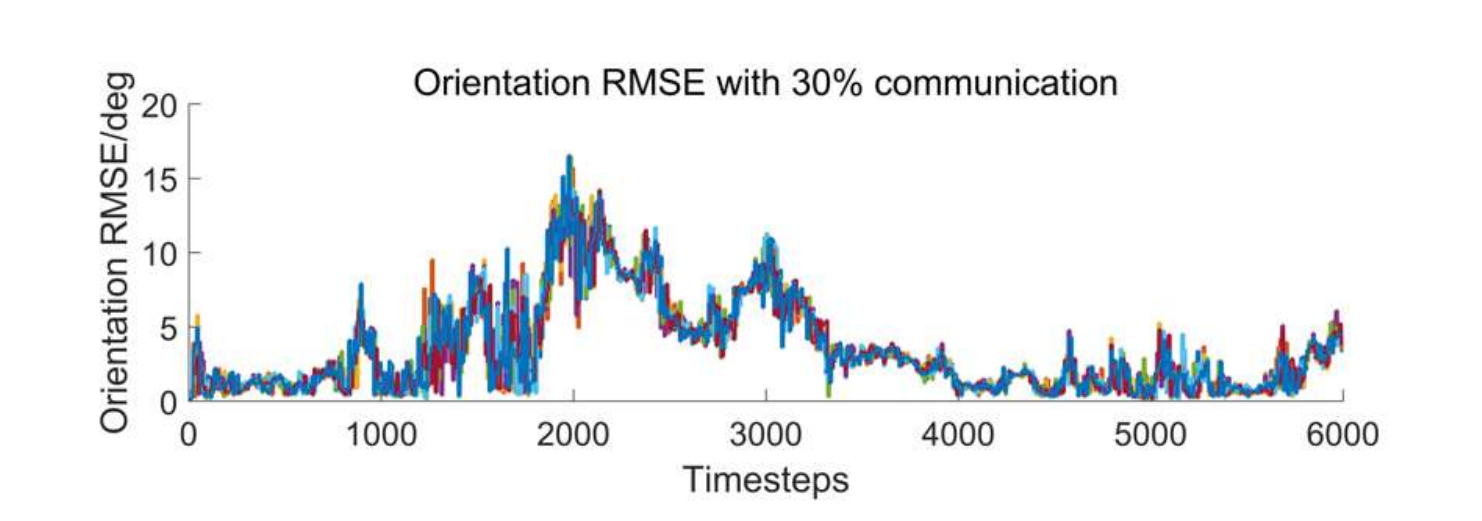}
		\caption{Orientation RMSE of each camera with $30\%$ communication rate}
	\end{subfigure}
        \caption{Orientation RMSE of each camera using the proposed algorithm with different communication rates.}
        \label{Fig_ORMSE}
\end{figure}

\begin{figure}[h!]
    \centering
    \includegraphics[width=0.48\textwidth]{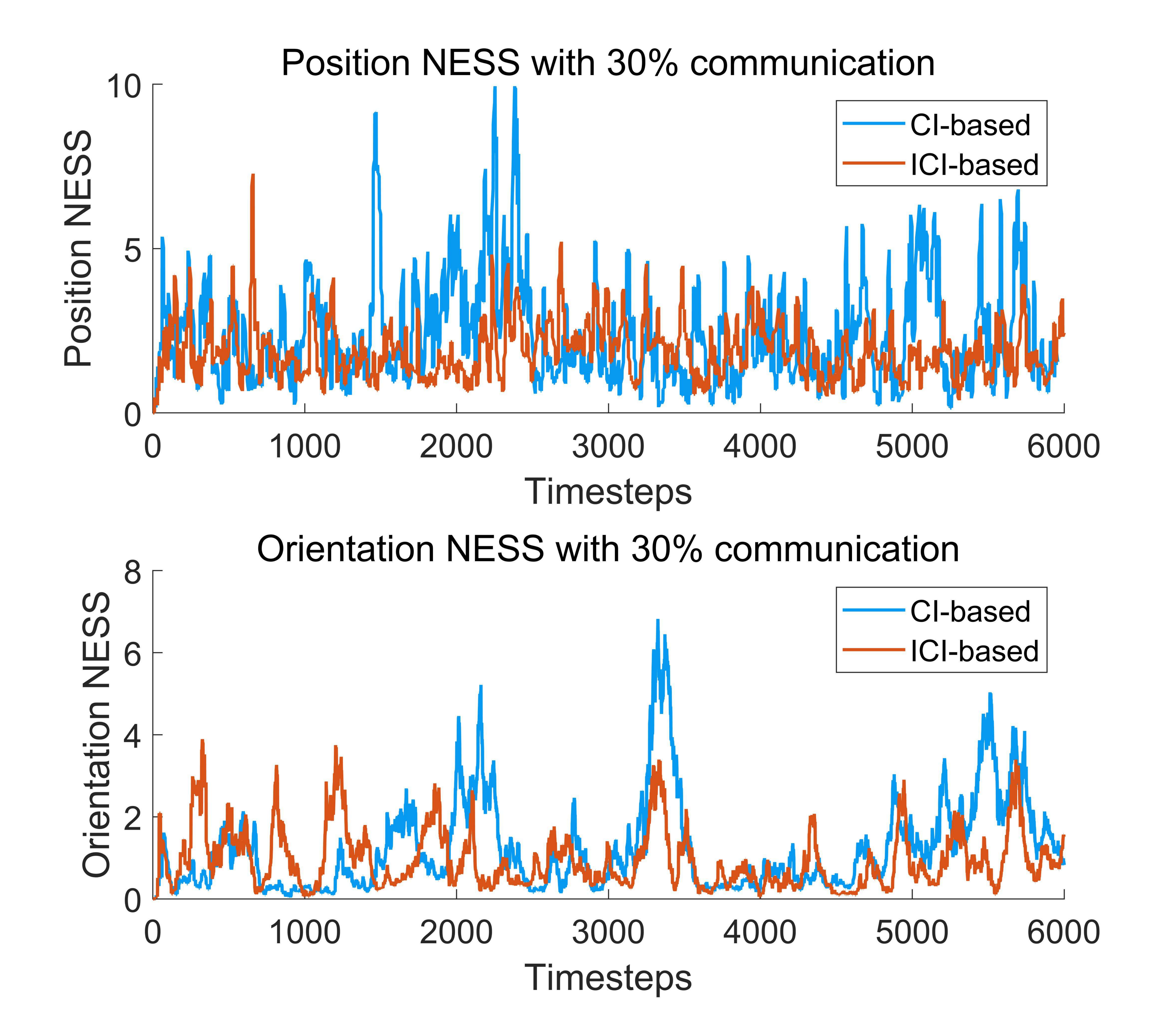}
    \caption{Comparison between the ICI-based and CI-based approach of averaged NEES with $30\%$ communication.}
    \label{fig6}
\end{figure}

We evaluate the performance of the proposed ICI-based tracking algorithm by comparing it with the centralized tracking approach and the CI-based approach \cite{Cras2009}. 
To illustrate the advantages of fusion performance, we test the tracking results with different communication rates across all the trails, as shown in table \ref{Tab1}. Obviously, without communication, or equivalently, when agents individually estimate the states of the target, there is no difference between CI and ICI. The performance is very poor due to the fact that the agents are blind for some time steps during testing. 
As soon as communication is established, the shared information leads to significantly improved fusion accuracy, and increasing the communication rate further enhances the estimation results, which shows the benefits of information fusion.
The performance of the proposed approach with $80\%$ communication rate is already comparable to that of the centralized method. It can also be observed that our algorithm outperforms the CI-based algorithm in estimating both position and orientation. Figure \ref{fig6} shows the averaged NEES result at $30\%$ communication rate. which illustrates the improvements of consistency for the proposed ICI-based fusion.

\begin{figure}[t]
	\centering
	\begin{subfigure}{0.46\textwidth}
		\centering
		\includegraphics[width=\textwidth]{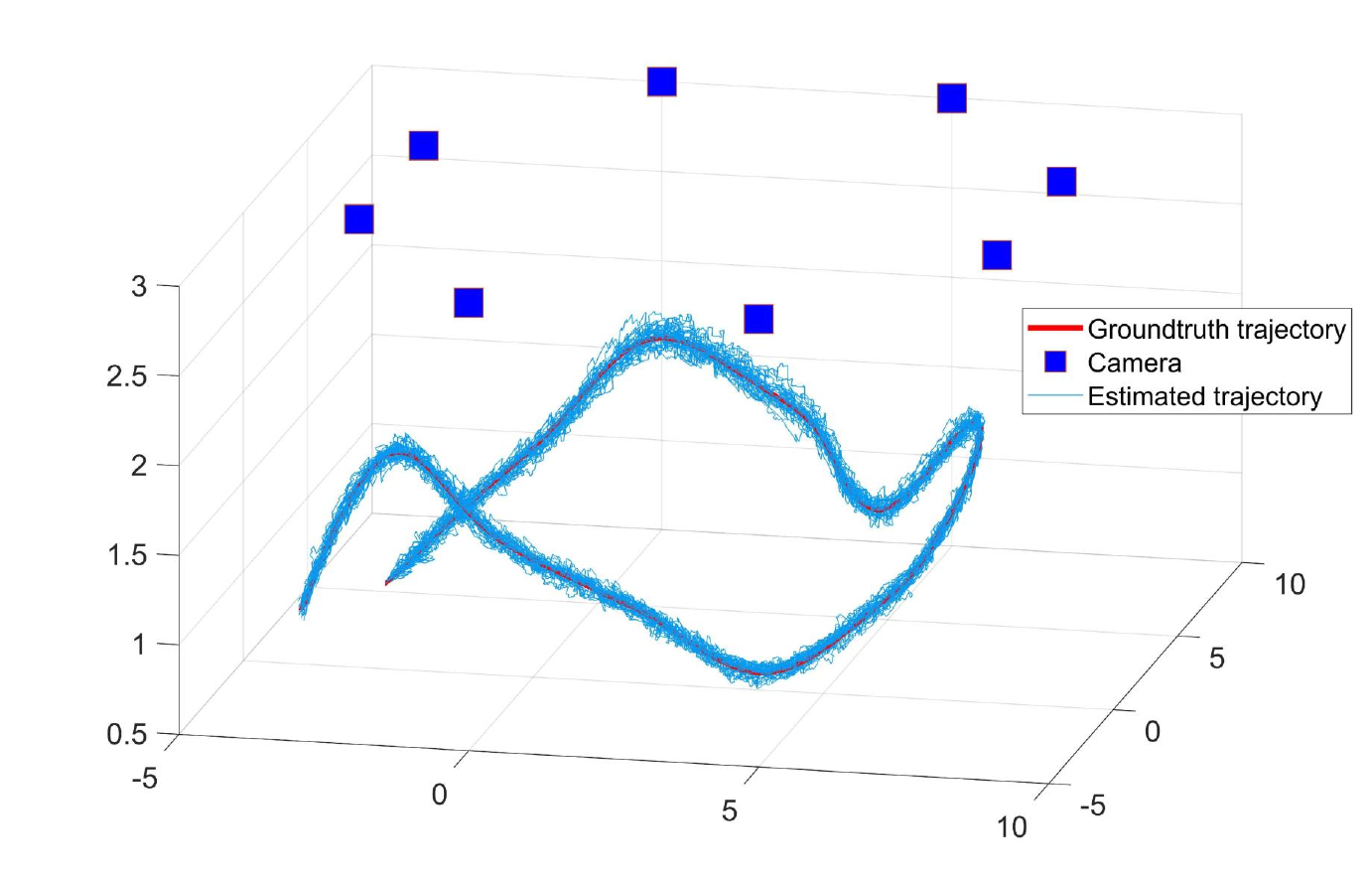}
		\caption{Estimation performance of trajectory (a)}
	\end{subfigure}
        \begin{subfigure}{0.46\textwidth}
		\centering
		\includegraphics[width=\textwidth]{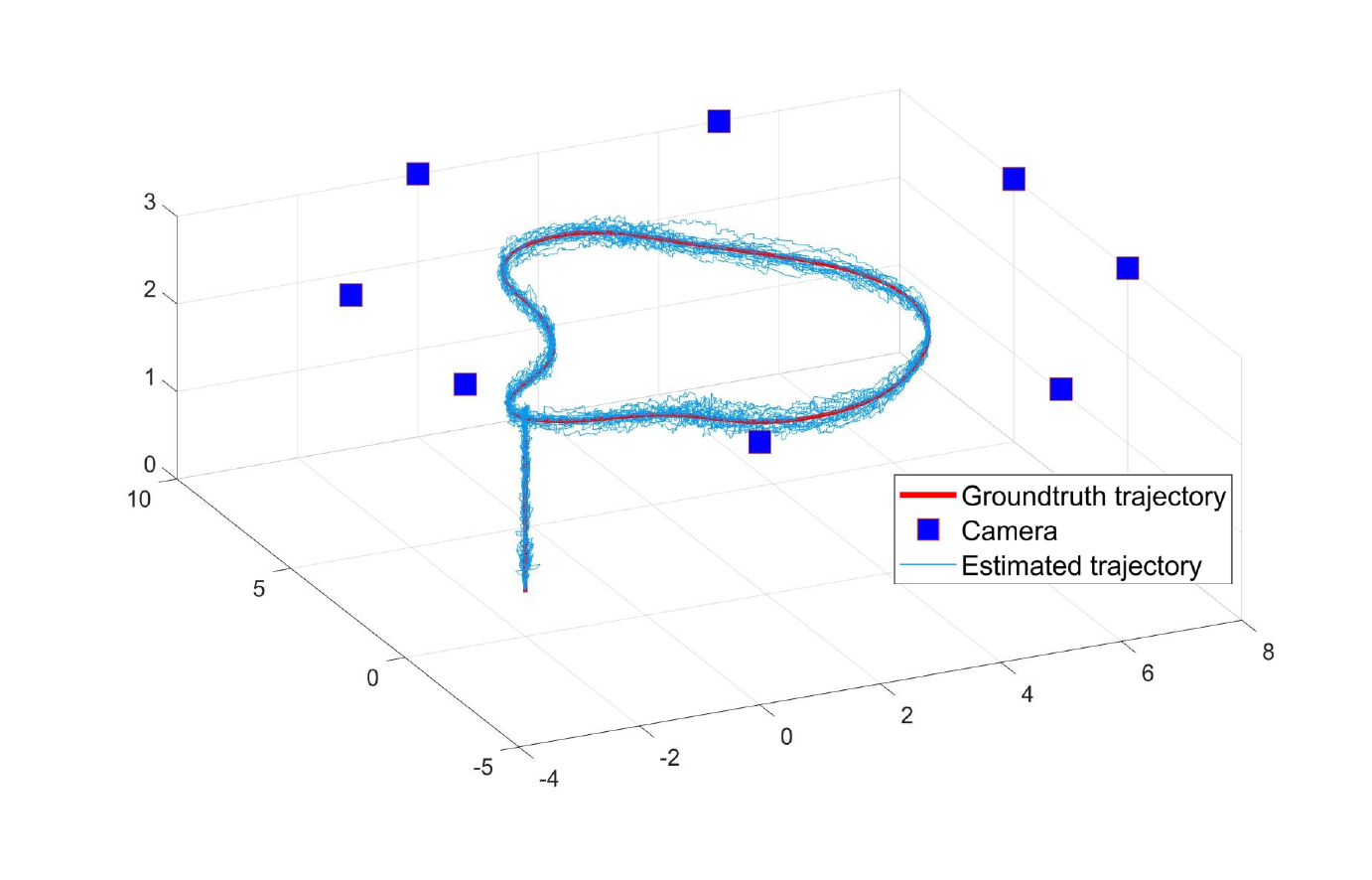}
		\caption{Estimation performance of trajectory (b)}
	\end{subfigure}
        \caption{Quadrotor's grountruth path and the tracking performance over 50 Monte-Carlo simulation with $30\%$ communication rate. We plot the estimated trajectories of the first 20 trails, which shows that the estimated trajectories are close to the groundtruth.}
        \label{Fig_Traj}
\end{figure}

To further show the performance of the individual cameras, we display the estimated trajectory of the first camera for both path $a$ and $b$ as shown in figure \ref{Fig_Traj}. Additionally, we show the position RMSE and the orientation RMSE of the path $a$ of each camera with different communication rates in figure \ref{Fig_PRMSE} and \ref{Fig_ORMSE}, that shows the 3-D moving target can be well-tracked by our algorithm. 


\section{Conclusion}
In this paper, we proposed a cooperative estimation framework by extending the ICI algorithm to augmented quaternion states, which is applicable for tracking a 3-D moving target over sensor networks. The proposed algorithm is fully distributed as it only uses its own and its neighbors' information, and is applicable to time-varying communication topology with blind agents. We validated the effectiveness of the proposed algorithm by Monte Carlo simulations which showed that the fused result is consistent and has comparable performance to centralized fusion when the communication probability among cameras is high.
Our future work will involve implementing the proposed algorithm on real robots for not only target estimation but also cooperative-localization and will extend the current framework to other filters that can deal with non-Gaussian noise.

\bibliographystyle{IEEEtran}
\bibliography{references}

\begin{thebibliography}{10}
\providecommand{\url}[1]{#1}
\csname url@samestyle\endcsname
\providecommand{\newblock}{\relax}
\providecommand{\bibinfo}[2]{#2}
\providecommand{\BIBentrySTDinterwordspacing}{\spaceskip=0pt\relax}
\providecommand{\BIBentryALTinterwordstretchfactor}{4}
\providecommand{\BIBentryALTinterwordspacing}{\spaceskip=\fontdimen2\font plus
\BIBentryALTinterwordstretchfactor\fontdimen3\font minus \fontdimen4\font\relax}
\providecommand{\BIBforeignlanguage}[2]{{%
\expandafter\ifx\csname l@#1\endcsname\relax
\typeout{** WARNING: IEEEtran.bst: No hyphenation pattern has been}%
\typeout{** loaded for the language `#1'. Using the pattern for}%
\typeout{** the default language instead.}%
\else
\language=\csname l@#1\endcsname
\fi
#2}}
\providecommand{\BIBdecl}{\relax}
\BIBdecl

\bibitem{Pzhu2021}
P.~Zhu and W.~Ren, ``Fully distributed joint localization and target tracking with mobile robot networks,'' \emph{IEEE Transactions on Control Systems Technology}, vol.~29, no.~4, pp. 1519--1532, 2021.

\bibitem{Yizhi2023}
Y.~Zhou, C.~Nowzari, and X.~Wang, ``Distributed multi-robot flocking based on acoustic doppler effect,'' in \emph{2023 62nd IEEE Conference on Decision and Control (CDC)}, 2023, pp. 6406--6411.

\bibitem{Zbu2022}
Y.~Wei, Z.~Wei, Y.~Rao, J.~Li, J.~Zhou, and J.~Lu, ``Lidar distillation: Bridging the beam-induced domain gap for 3d object detection,'' in \emph{Computer Vision -- ECCV 2022}, S.~Avidan, G.~Brostow, M.~Ciss{\'e}, G.~M. Farinella, and T.~Hassner, Eds.\hskip 1em plus 0.5em minus 0.4em\relax Cham: Springer Nature Switzerland, 2022, pp. 179--195.

\bibitem{Wang2009}
X.~Wang, S.~Wang, and D.~Bi, ``Distributed visual-target-surveillance system in wireless sensor networks,'' \emph{IEEE Transactions on Systems, Man, and Cybernetics, Part B (Cybernetics)}, vol.~39, no.~5, pp. 1134--1146, 2009.

\bibitem{Lig1997}
M.~Liggins, C.-Y. Chong, I.~Kadar, M.~Alford, V.~Vannicola, and S.~Thomopoulos, ``Distributed fusion architectures and algorithms for target tracking,'' \emph{Proceedings of the IEEE}, vol.~85, no.~1, pp. 95--107, 1997.

\bibitem{trawny2005}
N.~Trawny and S.~I. Roumeliotis, ``Indirect {Kalman} filter for 3d attitude estimation,'' \emph{University of Minnesota, Dept. of Comp. Sci. \& Eng., Tech. Rep}, vol.~2, p. 2005, 2005.

\bibitem{Eka2005}
E.~Kaplan and C.~Hegarty, \emph{Understanding GPS/GNSS: Principles and Applications, Third Edition}, 2017.

\bibitem{Hlink2014}
O.~Hlinka, O.~Slučiak, F.~Hlawatsch, and M.~Rupp, ``Distributed data fusion using iterative covariance intersection,'' in \emph{2014 IEEE International Conference on Acoustics, Speech and Signal Processing (ICASSP)}, 2014, pp. 1861--1865.

\bibitem{Pyi2023}
J.~Xu, P.~Zhu, Y.~Zhou, and W.~Ren, ``Distributed invariant extended kalman filter using lie groups: Algorithm and experiments,'' \emph{IEEE Transactions on Control Systems Technology}, vol.~31, no.~6, pp. 2777--2789, 2023.

\bibitem{Ara2001}
P.~Arambel, C.~Rago, and R.~Mehra, ``Covariance intersection algorithm for distributed spacecraft state estimation,'' in \emph{Proceedings of the 2001 American Control Conference. (Cat. No.01CH37148)}, vol.~6, 2001, pp. 4398--4403 vol.6.

\bibitem{Pzhu2020}
P.~Zhu and W.~Ren, ``Distributed kalman filter for 3-d moving object tracking over sensor networks,'' in \emph{2020 59th IEEE Conference on Decision and Control (CDC)}, 2020, pp. 2418--2423.

\bibitem{Liji2002}
L.~Chen, P.~Arambel, and R.~Mehra, ``Fusion under unknown correlation - covariance intersection as a special case,'' in \emph{Proceedings of the Fifth International Conference on Information Fusion. FUSION 2002. (IEEE Cat.No.02EX5997)}, vol.~2, 2002, pp. 905--912 vol.2.

\bibitem{Rein2015}
M.~Reinhardt, B.~Noack, P.~O. Arambel, and U.~D. Hanebeck, ``Minimum covariance bounds for the fusion under unknown correlations,'' \emph{IEEE Signal Processing Letters}, vol.~22, no.~9, pp. 1210--1214, 2015.

\bibitem{NOACK2017}
\BIBentryALTinterwordspacing
B.~Noack, J.~Sijs, M.~Reinhardt, and U.~D. Hanebeck, ``Decentralized data fusion with inverse covariance intersection,'' \emph{Automatica}, vol.~79, pp. 35--41, 2017. [Online]. Available: \url{https://www.sciencedirect.com/science/article/pii/S0005109817300298}
\BIBentrySTDinterwordspacing

\bibitem{Ajgl2020}
J.~Ajgl and O.~Straka, ``Inverse covariance intersection fusion of multiple estimates,'' in \emph{2020 IEEE 23rd International Conference on Information Fusion (FUSION)}, 2020, pp. 1--8.

\bibitem{Noa2019}
B.~Noack, U.~Orguner, and U.~D. Hanebeck, ``Nonlinear decentralized data fusion with generalized inverse covariance intersection,'' in \emph{2019 22th International Conference on Information Fusion (FUSION)}, 2019, pp. 1--7.

\bibitem{Sun2023}
T.~Sun and M.~Xin, ``Inverse-covariance-intersection-based distributed estimation and application in wireless sensor network,'' \emph{IEEE Transactions on Industrial Informatics}, vol.~19, no.~10, pp. 10\,079--10\,090, 2023.

\bibitem{SunT2023}
S.~Tao and X.~Ming, ``Distributed estimation with iterative inverse covariance intersection,'' vol.~70, no.~4, 2023, pp. 1645--1649.

\bibitem{Qi2023}
\BIBentryALTinterwordspacing
L.~Qi, X.~Wang, and C.~Cui, ``Augmented quaternion and augmented unit quaternion optimization,'' 2023. [Online]. Available: \url{https://api.semanticscholar.org/CorpusID:257219095}
\BIBentrySTDinterwordspacing

\bibitem{Juli1997}
S.~Julier and J.~Uhlmann, ``A non-divergent estimation algorithm in the presence of unknown correlations,'' in \emph{Proceedings of the 1997 American Control Conference (Cat. No.97CH36041)}, vol.~4, 1997, pp. 2369--2373 vol.4.

\bibitem{Cras2009}
C.~K.~N. John L.~Crassidis, Yang~Cheng and A.~M. Fosbury, ``Decentralized attitude estimation using a quaternion covariance intersection approach,'' \emph{The Journal of the Astronautical Sciences}, vol.~11, pp. 113--128, 2009.

\bibitem{Huang2011}
G.~P. Huang, A.~I. Mourikis, and S.~I. Roumeliotis, ``An observability-constrained sliding window filter for slam,'' in \emph{2011 IEEE/RSJ International Conference on Intelligent Robots and Systems}, 2011, pp. 65--72.

\end{thebibliography}

\begin{appendix}
\subsection{IMU propagation model}\label{IMU_mdl}
We use a 6-axis on-board IMU for the propagation. Suppose that the IMU's frame is exactly the target's local frame $L$. The IMU provides measurements of the translational accelerations ${\bm a}_m$ and angular rates $\omega_m$ acting on the system with respect to the IMU's local frame $ L$
\begin{align}
    \bm\omega_m(t)&={^L{\bm\omega}}(t)+\mathbf{n}_\omega(t)\nonumber\\
    {\mathbf a}_m(t)&={^L\mathbf a}(t)+\mathbf{n}_a(t)\nonumber+^L_G{\mathbf R}(t)\bm g
\end{align}
where ${^L{\bm\omega}}(t)$ and ${^L\mathbf a}(t)$ is the true angular rate and translational acceleration with respect to the IMU's local frame $L$, $\bm n_a$, $\bm n_\omega$ are IMU noises with Gaussian distribution.
According to IMU dynamics, the continuous target's motion model can be represented by \cite{trawny2005}
\begin{align}\label{IMU_dyn}
^L_G{\dot{\mathbf q}(t)}&=\frac{1}{2}
\begin{bmatrix}
-\lfloor {^L{\bm\omega}}(t)\times \rfloor&{^L{\bm\omega}}(t)\\
-{^L\omega}(t)&0
\end{bmatrix}{^L_G{\mathbf q}}(t)\nonumber\\
^G {\dot{\mathbf p}}(t)&={^G \mathbf v}(t)\nonumber\\
^G {\dot{\mathbf v}}(t)&=^L_G{\mathbf R}(t)^\top{^G \mathbf a}(t)
\end{align}
Recall the definition of error state from equation \eqref{err_state}. Linearizing equation \eqref{IMU_dyn} yields the corresponding error state dynamics obtained by each agent
\begin{align}\label{IMU_err_dyn}
\delta {\mathbf {\dot x}_i}(t)=F_i\delta {\mathbf {x}_i}(t)+G_i \mathbf r_i(t)
\end{align}
where 
\begin{align}
F_i&=
\begin{bmatrix}
-\lfloor \bm \omega_m \times\rfloor&\mathbf 0_{3\times 3}&\mathbf 0_{3\times 3}\\
\mathbf 0_{3\times 3}&\mathbf 0_{3\times 3}& \mathbf I_{3\times 3}\\
\lfloor ^L_G{\mathbf R}^\top \mathbf a_m \times\rfloor&\mathbf 0_{3\times 3}&\mathbf 0_{3\times 3}
\end{bmatrix}\\
\mathbf G_i&=
\begin{bmatrix}
-\mathbf I_{3\times 3}&\mathbf 0_{3\times3}\\
\mathbf 0_{3\times3}&\mathbf 0_{3\times3}\\
\mathbf 0_{3\times3}&-^L_G\mathbf{R}^\top
\end{bmatrix}
\end{align}
In the simulation, we discrete the dynamics \eqref{IMU_err_dyn} for propagation. The noise of the linear acceleration and angular velocity is set to be white Gaussian noise with standard deviations $0.02 m/{s^2}$ and $0.03 rad/s$ for the continuous IMU model.

\subsection{Camera measurement model}
Suppose that the camera is fixed in the global frame with rotation $^C_G\mathbf R$ from the global frame $\mathbf G$ to the camera frame $C$, and position $^G \mathbf p_c$. The standard pinhole camera model used to measure a 3-D moving target with measurement noise can be represented as
\begin{align}\label{cam_mdl}
z^k&=h_p^k(^C \mathbf p^k)+\mathbf w^k=\mathbf h_p ({^C_G\mathbf R}(^G \mathbf p^k-{^G \mathbf p_c}))=
\begin{bmatrix}
\frac{^c x}{^c z}\\\frac{^c y}{^c z}
\end{bmatrix}+\mathbf w^k
\end{align}
where ${^C \mathbf p^k}=\begin{bmatrix}^c x&^c y&^c z\end{bmatrix}^\top$ denotes the target position with respect to the camera frame, and $\mathbf h_p: \mathbb R^{3\times 1} \to\mathbb R^{2\times 1}$ is defined as the projection function given as
\begin{align}
h_p(\begin{bmatrix}
^c x&^c y &^c z
\end{bmatrix}^\top)=\begin{bmatrix}\frac{^c x}{^c z}\\\frac{^c y}{^c z}\end{bmatrix}
\end{align}
which project a 3-D target point in the camera frame into normalized image plane producing the 2-D pixel values. The state Jacobian of $h_p$ can be computed as
\begin{align}
H_p=\frac{\partial h_p({^C \mathbf p}^k)}{\partial{^C \mathbf p}^k}=
\begin{bmatrix}
\frac{1}{^C z}&0&-\frac{^C x}{{^C z}^2}\\
0&\frac{1}{^C z}&-\frac{^C y}{{^C z}^2}
\end{bmatrix}
\end{align}
and the Jacobian of \eqref{cam_mdl} evaluated at $^C \mathbf p^k$ can be computed as
\begin{align}\label{eq_def_Jac}
\mathbf H=H_p(^C \mathbf p^k){^C_G \mathbf R}
\begin{bmatrix}
\mathbf 0_{3\times 3}&\mathbf{I}_{3\times 3}&\mathbf 0_{3\times 3}
\end{bmatrix}
\end{align}

\end{appendix}

\addtolength{\textheight}{-12cm}   
\end{document}